\documentclass[12pt]{article}
\usepackage[left=2.5cm,top=2.50cm,right=2.5cm,bottom=2.50cm]{geometry}
\usepackage{mathrsfs}
\usepackage{amsmath,amssymb,latexsym,color,cancel,graphicx,bbm,colortbl}
\usepackage[english]{babel}
\usepackage[latin1]{inputenc}
\usepackage{ragged2e}
\usepackage{cite}
\DeclareMathAlphabet{\mathpzc}{OT1}{pzc}{m}{it}
\begin{document}
\date{}

\title{Algebraic solution and coherent states for the Dirac oscillator interacting with the Aharonov-Casher system in the cosmic string background}
\author{M. Salazar-Ram\'irez$^{a}$\footnote{{\it E-mail address:} escomphysics@gmail.com}, D. Ojeda-Guill\'en$^{a}$, \\ R. D. Mota$^{b}$, J. A.  Mart\'inez-Nuño$^{a}$, M. R. Cordero-L\'opez$^{a}$} \maketitle

\begin{minipage}{0.9\textwidth}
\small $^{a}$ Escuela Superior de C\'omputo, Instituto Polit\'ecnico Nacional,
Av. Juan de Dios B\'atiz esq. Av. Miguel Oth\'on de Mendiz\'abal, Col. Lindavista,
Delegaci\'on Gustavo A. Madero, C.P. 07738, Ciudad de M\'exico, M\'exico.\\

\small $^{b}$ Escuela Superior de Ingenier{\'i}a Mec\'anica y El\'ectrica, Unidad Culhuac\'an,
Instituto Polit\'ecnico Nacional, Av. Santa Ana No. 1000, Col. San Francisco Culhuac\'an, Delegaci\'on Coyoac\'an, C.P. 04430, Ciudad de M\'exico, Mexico.\\
\end{minipage}

\begin{abstract}

We introduce an $SU(1,1)$ algebraic approach to study the $(2+1)$-Dirac oscillator in the presence of the Aharonov-Casher effect coupled to an external electromagnetic field in the Minkowski spacetime and the cosmic string spacetime. This approach is based on a quantum mechanics factorization method that allows us to obtain the $su(1,1)$ algebra generators, the energy spectrum and the eigenfunctions. We obtain the coherent states and their temporal evolution for each spinor component of this problem. Finally, for these problems, we calculate some matrix elements and the Schr\"odinger uncertainty relationship for two general $SU(1,1)$ operators.

\end{abstract}


\section{Introduction}

It is known from the literature that when a system undergoes an evolution and returns to its original state after a certain time, it is a cyclic evolution. When a system is classical it is impossible to detect such an evolution, but in quantum mechanics it is possible because the wave function in a quantum system retains a memory of its motion in the form of a geometric phase factor, which can be measured by interfering the wave function with another coherent wave function\cite{Anan,Cohen2}. Thus, it is possible to discern whether or not the system has undergone an evolution. Anandan et. al. in their article defined the geometric phase factors as ``signatures" of quantum motion \cite{Anan}. Also, in that reference they pointed out that the adjective geometric in the phase factor simply emphasizes that such factors depend only on the loop in the quantum-mechanical state space (the set of rays of Hilbert space), sometimes called the projective Hilbert space.

Although the geometric phase is attributed to quantum mechanics, it has been applied to many areas in physics, such as condensed-matter, optics, fluid mechanics, gravity and cosmology, and particle physics \cite{Cohen2}. The two classical examples of geometric phases are the Aharonov-Bohm (AB) \cite{Aha} and Aharonov-Casher (AC) \cite{AhaC}. In fact, these are the simplest examples illustrating the Berry phase concept.

In $1959$, AB published their work on the importance of electromagnetic potentials in quantum theory. This problem is now known as the AB effect, and consists of an electron beam that is emitted from a source on a plate with two slits (double slit experiment) that passes around an infinitely long solenoid. It is expected that these electrons do not experiment any type of electromagnetic force, because the magnetic field is confined in the solenoid. However, a phase change is observed in the electron beam, that is, the electron beam is influenced by the electromagnetic field even when it does not come into contact with this field.

The AC effect corresponds to the appearance of a geometric quantum phase in the wave function of a neutral particle with magnetic dipole moment when the particle moves in a plane under the influence of an electric field produced by a uniformly charged infinitely long filament perpendicular to the plane. Therefore when the particle moves around the filament, the wave function will acquire a phase equal to $\mu\lambda$, where $\mu$ is the magnetic dipole moment of the particle and $\lambda$ is the linear electric charge density of the filament\cite{He}, i.e.
 \begin{equation}
\Phi_{AC}=\oint\left(\overrightarrow{\mu}\times\overrightarrow{E}\right)\cdot d\overrightarrow{r}=2\pi s\mu\lambda,
\end{equation}
where $s$ represents the spin degrees of freedom, i.e. different components acquire phases with different signs. The AC effect is based on the fact that the term $\left(\overrightarrow{\mu}\times\overrightarrow{E}\right)=s\Phi_{AC}/2\phi\rho\widehat{\varphi}$  plays the role of an effective vector potential, and it is the responsible for the emergence of a phase change in the wave function of the neutral particle.

It has been possible to experimentally measure the AC phase with different practical approaches \cite{Spav,Cimmi}. K. Sangster et al. consider the AC effect as a topological effect regarding the relative orientation of the electric field and the magnetic moments \cite{Sang}. Despite the auxiliary role of electromagnetic potentials in classical electromagnetism, the topological quantum AB and AC effects play a relevant physical role in quantum theory \cite{Spav}. It has been argued that the gravitational field of the Earth introduces a spin independent phase shift. Therefore, it was used unpolarized neutrons to find the AC predicted phase shift change for the magnetic dipole diffracting around a charged electrode for the case of thermal neutrons. To this end it was used a neutron interferometer containing a $30-kV/mm$ vacuum electrode system and it was found a phase shift of $2.19\pm0.52$  $mrad$ \cite{Cimmi}.

Mirza and Zarei obtained the corrections to the topological phase of the AC effect in a noncommutative space, where the linear spectrum does not depend on the relativistic nature of the dipoles. Also, these authors obtained a generalized formula for holonomy \cite{Mirza}.

The study of the topological phases in the quantum dynamics of a single particle that moves freely in a multiply connected space-times in a variety of physical systems has aroused great interest. In recent years, the study of the Dirac oscillator interacting with topological defects has been widely studied from different points of view \cite{RRS,Fur,Bakkee}. In Ref. \cite{Fur} the influence of the AC effect on the Dirac oscillator (DO) was studied and the energy spectrum and the eigenfunctions for bounded states were obtained. Also, in Ref. \cite{Fur} it was showed that the relativistic energy levels depend on the AC geometric phase in the Minkowski spacetime (Ms), the cosmic string spacetime (Css) and the cosmic dislocation spacetime. In Ref. \cite{Bakke2}, Barbosa et. al. studied the AC phase on a Coulomb type potential, showing that the energy levels depend on the AC geometric phase. Moreover, these authors studied the Landau-Aharonov-Casher (LAC) phase corresponding to a Coulomb type potential, obtaining the bound states solutions for the Schr\"odinger-Pauli equation \cite{Bakke2}.

It was Glauber who first noticed that the simplest way to ensure the fully coherence for the quantized fields is to require the quantum state of the field to be an eigenvector of the bosonic annihilation operator with complex eigenvalue, which is a solution (written in complex form) of the corresponding Maxwell equations \cite{Glau}. Therefore, the fully coherent Glauber states of the quantized electromagnetic radiation are directly associated with conventional electromagnetic theory. In fact, Glauber defined the coherent states as the eigenstates of the annihilation operator of the harmonic oscillator which are related to the Heisenberg-Weyl group. A more detailed review of the main properties of coherent states can be found in Ref. \cite{RosasO}.

Talking about the origin of coherent states it is inevitable to mention to Schr\"odinger \cite{Sch}, due to the coincidence between his wave packet and the $x_1$ -quadrature representation of the Glauber's coherent states. Schr\"odinger found this wave packet when he was looking for a connection between the classical world and the harmonic oscillator quantum states. However, it is important to mention that the Schr\"odinger wave packet and the Glauber coherent states were obtained by different approaches \cite{RosasO}.

The relationship between coherent states and the $SU(1,1)$ group was given by Barut and Girardello. They introduced the $SU(1,1)$ coherent states as the eigenstates of the annihilation operator of this group \cite{Barut}. Perelomov defined his coherent states as those resulted from the action of a Lie group on a fiducial state, which is defined by its invariance under the action of the basis element that parameterizes the representation \cite{Perelomov2,Gilmore2,Puri}.

Thus, because of the previous background we are motivated to study the Dirac equation problem for the Dirac oscillator including the Aharonov-Casher effect coupled to an external electromagnetic field in the Ms and in the Css from an algebraic approach. By using the Schr\"odinger factorization we construct a set of radial operators which close the $su(1,1)$ Lie algebra. The relevant results obtained in this paper are based on this $SU(1,1)$ symmetry.

The paper is organized as follows. In Section $2$, we give a review of the DO and the AC effect in the Ms to obtain the uncoupled differential equations for each spinor component. In Section $3$, we apply the Schr\"odinger factorization to construct three radial operators (which depend on the magnetic field strength, the AC frequency and the AC phase) that close the $su(1,1)$ Lie algebra. The theory of unitary representations let us to obtain the energy spectrum and the eigenfunctions in an algebraic way. Moreover, we are able to construct the Perelomov coherent states and their time evolution for the radial equations by using the Sturmian basis of the $su(1,1)$ Lie algebra. In Sections $4$ and $5$, we obtain analogous results for the DO and the AC effect in the Css. In Section $6$, we obtain some matrix elements for the radial eigenfunctions. Also, we show that the Schr\"odinger uncertainty relationship (\emph{Sur}) evaluated in the Perelomov coherent states holds for two general $SU(1,1)$ operators in both, the Ms and the Css. Finally, we give some concluding remarks.

\section{The DO and the AC effect}

The metric tensor in the Ms for the DO in the presence of the AC effect  with an external electromagnetic field is defined by the line element\cite {RRS}
 \begin{equation}
ds^2=c^2dt^2-d\rho^2-\rho^2d\theta^2,
\end{equation}
where $|t|<\infty$, $\theta\in[0,2\pi]$ is the azimuthal coordinate and $c$ the speed of light\cite{GREINER}.

The electric fields for the AC effect  are $\boldsymbol{E_1}$ and $\boldsymbol{E_2}$. $\boldsymbol{E_1}$ is generated by an infinitely long wire and uniformly charged, located along the $z$-axis, perpendicular to the polar plane, and $\boldsymbol{E_2}$ is produced in the inner region of a uniformly charged non-conducting cylinder of length $L$ and radius $R$. Explicitly,
\begin{align}\label{CEle1}
\boldsymbol{E_1}=&\frac{2\lambda_1}{\rho}\widehat{e}_\rho, \hspace{0.5cm} \nabla\cdot\boldsymbol{E_1}=2\lambda_1\frac{\delta(\rho)}{\rho}, \hspace {0.5cm} \left(\rho=\sqrt{x^2+y^2}\right),\\\label{CEle2}
\boldsymbol{E_2}=&\frac{\lambda_2\rho}{2}\widehat{e}_\rho, \hspace{0.5cm} \nabla\cdot\boldsymbol{E_2}=\lambda_2, \hspace{0.5cm} \frac{\partial \boldsymbol{E_2}}{\partial t}=0, \hspace{0.5cm} \nabla\times\boldsymbol{E_2}=0,
\end{align}
where $\widehat{e}_\rho$ is a unit vector on the radial direction, $\lambda_1=\lambda_0/4\pi\epsilon_0, \lambda_2=\chi/\epsilon _0$, $\epsilon_0$ is the electric vacuum permittivity, $\lambda_0>0$ is the electric charge linear density of the wire, $\rho>0$ is the radial coordinate and $\chi=(Q/\pi R^2L)>0$ $(R\ll L)$ is the electric charge volumetric density of the cylinder. Note that $\nabla\cdot\boldsymbol{E_1}$ gives rise to a two-dimensional $\delta$ function potential at origin, which allows at least one bound state \cite{Belich}. The magnetic field $\boldsymbol{B}$ originated by an infinitely long solenoid is expressed by the mathematical equations
\begin{equation}\label{CMag}
\boldsymbol{B}=B\widehat{e}_z, \hspace{0.5 cm} \nabla\cdot \boldsymbol{B}=0, \hspace{0.5 cm} \frac{\partial \boldsymbol{B}}{\partial t}=0, \hspace{0.5cm} \nabla\times\boldsymbol{B}=\mu_o\boldsymbol{J}.
\end{equation}

The covariant Dirac equation for a neutral fermion with electric dipolar momentum $\mu$ is \cite{GREINER}
\begin{equation}\label{ECUD2}
\left[i\gamma^{a}\partial_a+\frac{\mu}{2}\sigma^{ab}F_{ab}-m_0\right]\Psi(t,\textbf{r})=0,\hspace{2.0 cm} (a,b=0,1,2),
\end{equation}
with the matrices $\gamma^{a}$ satisfying the anticommutation relations of the Clifford Algebra $\{\gamma^a, \gamma^b\}=2\eta^{ab}$, being $\eta^{ab}$ the Minkowski metric tensor, $\sigma^{ab}=\frac{1}{2}\left[\gamma^a, \gamma^b\right]$ and $F_{ab}$ the electromagnetic field tensor. If we substitute $\frac{1}{2}\sigma^{ab}F_{ab}=i\boldsymbol{\alpha}\cdot\mathbf{E}- \boldsymbol{\Sigma}\cdot\boldsymbol{B}$ into equation (\ref{ECUD2}) we obtain
\begin{equation}
\left[i\gamma^0\partial_0-\boldsymbol{\gamma}\cdot\left(\boldsymbol{p}-im_0\omega\beta\boldsymbol{r}\right)+\mu\left(i\boldsymbol{\alpha}\cdot \boldsymbol{E}- \boldsymbol{\Sigma}\cdot \boldsymbol{B}\right) -m_0\right]\Psi(t,\boldsymbol{r})=0,
\end{equation}
where the matrices in the standard Dirac representation are given as
\begin{equation}
\boldsymbol{\alpha}=\widehat{\beta}\boldsymbol{\gamma}=\begin{pmatrix}
0 & \boldsymbol{\sigma} \\
\boldsymbol{\sigma} & 0
\end{pmatrix}
, \hspace{0.5cm} \boldsymbol{\Sigma}=\begin{pmatrix}
\boldsymbol{\sigma} & 0 \\
 0 & \boldsymbol{\sigma}
\end{pmatrix}.
\end{equation}
After replacing equations (\ref{CEle1}), (\ref{CEle2}) and (\ref{CMag}) in equation (\ref{ECUD2}) we get the following Dirac equation
\begin{equation}\label{ARA-1}
\left[i\gamma^0\partial_0+i\gamma^\rho\left(\partial_\rho+\gamma^0\left(m_0\bar{\omega}\rho-\frac{s\Phi_{AC}}{\pi\rho}\right)\right)+i\frac{\gamma^\theta\partial_\theta}{\rho}-\Sigma^z\mu{B}-m_0\right]\Psi(t,\rho,\theta)=0,
\end{equation}
where $\gamma^0=\beta, \gamma^\rho=\boldsymbol{\gamma}\cdot\widehat{e}_{\rho}=\gamma^1\cos\theta+\gamma^2\sin\theta, \gamma^\theta=\boldsymbol{\gamma}\cdot\widehat{e}_\theta=-\gamma^1\sin\theta+\gamma^2\cos\theta, \Sigma^z=\boldsymbol{\Sigma}\cdot\widehat{e}_z, \bar{\omega}=\left(\omega- \frac{\omega_{AC}}{2}\right)\geq0, \omega_{AC}=\frac{\mu\lambda_2}{m_0}$ is the AC frequency of the fermion and $\Phi_{AC}=2s\pi\mu\lambda_1$  is the AC phase \cite{mer}. The term $s=\pm 1$ represent the projections of the magnetic dipole moment of the fermion along on the $z$-axis \cite{Fur,He,KS}.

In order to reduce the Dirac matrices $\gamma^\rho$ and $\gamma^\theta$ to fixed matrices $\gamma^1$ and $\gamma^2$ we apply to them the following similarity transformation
\begin{equation}
U^{-1}(\theta)\gamma^\rho U(\theta)=\gamma^1, \hspace{0.5cm} U^{-1}(\theta)\gamma^\theta U(\theta)=\gamma^2,
\end{equation}
where $U(\theta)=e^{-\frac{i\theta\alpha_3}{2}}$. Also, in a $(2+1)$- dimensional Ms the Dirac matrices $\boldsymbol{\gamma}=(\gamma^1,\gamma^2)=(-\gamma_1,-\gamma_2)$, $\gamma^0=\Sigma^z$ are given in terms of the Pauli matrices as $\gamma_1=\sigma_3\sigma_1$, $\gamma_2=\sigma_3\sigma_2$ and $\gamma^0=\Sigma^z=\sigma_3$. Thus, equation (\ref{ARA-1}) can be written as
\begin{equation}\label{ARA-2}
\left[\sigma_1\left(i\partial_\rho+i\sigma_3\left(m_0\bar{\omega}\rho-\frac{s\Phi_{AC}}{\pi\rho}\right)\right)+\sigma_2\left(\frac{i\partial_\theta}{\rho}+\frac{\sigma_3}{2\rho}\right)+\mu{B}+m_0\sigma_3-i\partial_0\right]\psi(t,\rho,\theta)=0,
\end{equation}
where $\psi(t,\rho,\theta)\equiv U^{-1}(\theta)\Psi(t,\rho, \theta)$ and the Dirac spinor is given by \cite{GREINER}
\begin{equation}
 \Psi(t,\rho,\theta)=\frac{e^{i\left( m_l\theta-Et\right)}}{\sqrt{2\pi}}\begin{bmatrix}
\varphi_+(\rho)\\
i\varphi_-(\rho)
\end{bmatrix}, \hspace{0.5 cm} \left(m_l=\pm 1/2,\pm3/2,...\right).
\end{equation}

Therefore, we can write equation (\ref{ARA-2}) as the two first-order coupled differential equations
\begin{align}\label{acop1}
&\left(m_0+\mu{B}-E\right)\varphi_+(\rho)=\left[\frac{d}{d\rho}-m_0\bar{\omega}\rho+\frac{1}{\rho}\left(m_l+\frac{s\Phi_{AC}}{\pi}+\frac{1}{2}\right)\right]\varphi_-(\rho),\\ &\left(m_0-\mu{B}+E\right)\varphi_-(\rho)=\left[\frac{d}{d\rho}+m_0\bar{\omega}\rho-\frac{1}{\rho}\left(m_l+\frac{s\Phi_{AC}}{\pi}-\frac{1}{2}\right)\right]\varphi_+(\rho).
\end{align}

By combining these equations, we obtain the uncoupled differential equations for the radial components
\begin{equation}\label{secondAC}
\left[\frac{d^2}{d\rho^2}+\frac{1}{\rho}\frac{d}{d\rho}-\frac{\Gamma_s^2}{\rho^2}-m_0^2\bar{\omega}^2\rho^2+E_s\right]\varphi_s\left(\rho \right)=0,\hspace{0.5cm} \left(s=\pm 1\right),
\end{equation}
with
\begin{align}\label{cmvar}
\Gamma_s\equiv & \mu_l+\frac{s\Phi_{AC}}{\pi}-\frac{s}{2},\\\label{xi}
E_s\equiv &\left(\mu B-E\right)^2-m_0^2+2m_0\bar{\omega}\Gamma_s+2sm_0\bar{\omega}.
\end{align}

In these expressions, $E$ is the relativistic total energy of the fermion and $\mu_l$ is the orbital magnetic quantum number. It is important to note that, since the AC geometric phase $\Phi_{AC}$ appears in the centrifugal term, this will directly affect $\mu_l$. However, since $\Gamma_s$ also appears in equation (\ref{xi}), the phase $\Phi_{AC}$ will also modify the energy spectrum. The effect of this phase $\Phi_{AC}$ in this energy spectrum will be discussed later. Further, the equation (\ref{secondAC}) is mathematically the same radial equation of the three-dimensional harmonic oscillator. Thus, we can anticipate that this problem is completely solvable and, in order to obtain it solution, we can apply a suitable algebraic approach like the Schr\"odinger factorization.

\section{Algebraic solution for the DO and the AC effect}

In this Section, we shall construct the $su(1,1)$ algebra generators of the equation (\ref{secondAC}) by using the Schr\"odinger factorization \cite{MSR1,MSR2,Schro,Schro1,Schro2,Inf}. In order to solve this general equation we must eliminate the first derivative by the change variable $\varphi_s\left(\rho \right)=\frac{1}{\sqrt{\rho}}\Psi_s\left(\rho \right)$. Thus, equation (\ref{secondAC}) is written as
\begin{equation}\label{rocuad}
\left(-\rho^2\frac{d^2}{d\rho^2}+m^2\overline{\omega}^2\rho^4-E_s\rho^2\right)\Psi_s=\left(\frac{1}{4}-\Gamma_s^2\right)\Psi_s,
\end{equation}
from which we construct the $su(1, 1)$ algebra generators by applying the Schr\"odinger factorization. Thus, we propose
\begin{equation}
\left(\rho\frac{d}{d\rho}+\Xi\rho^2+\Omega\right)\left(-\rho\frac{d}{d\rho}+\Lambda\rho^2+\Pi\right)\Psi_s=\Theta\Psi_s.
\end{equation}
If the last expression is expanded and compared with equation (\ref{rocuad}), we find that the constants $\Xi,\Omega,\Lambda$ and $\Theta$ are
\begin{equation}
\Xi=m\overline{\omega}, \hspace{0.5cm} \Pi=\Omega+1=-\frac{E_s}{2m\omega}-\frac{1}{2}, \hspace{0.5cm} \Theta=\left(\frac{E_s}{2m\overline{\omega}+1}\right)^2-\Gamma_s^2.
\end{equation}
Thus, equation (\ref{rocuad}) can be factorized as
\begin{equation}
\left[\mathbb{A}_-^s-1\right]\mathbb{A}_+^s=\frac{1}{4}\left[\left(\frac{E_s}{2m_0|\bar{\omega}|}+1\right)^2-\Gamma_s^2\right],
\end{equation}
where $\mathbb{A}_\mp^s$ are the Schr\"odinger operators explicitly given by
\begin{equation}\label{ED-COM}
\mathbb{A}_\mp^s=\frac{1}{2}\left[\pm\rho\frac{d}{d\rho}+m_0|\bar{\omega}|\rho^2-\frac{1}{2m_0\bar{\omega}}\left[\left(\mu B-E\right)^2-m_0^2+2m_0|\bar{\omega}|\Gamma_s+2m_0|\bar{\omega}|\right]+s \right].
\end{equation}
From these expressions we define the pair of operators
\begin{align}\label{OPSCHAC1}
\mathbb{B}_{\pm(1)}^s=&\frac{1}{2}\left[\mp\rho\frac{d}{d\rho}+m_0|\bar{\omega}|\rho^2-\frac{m_0^2-2m_0|\bar{\omega}|\left(\mu_l+\frac{s\Phi_{AC}}{\pi}-\frac{s}{2}\right)}{2m_0|\bar{\omega}|}\right]+\frac{1-s}{2}-\mathbb{B}_3^s,\\\label{OPSCHAC2}
\mathbb{B}_{\pm(2)}^s=&\frac{1}{2}\left[\mp\rho\frac{d}{d\rho}+m_0|\bar{\omega}|\rho^2-\frac{m_0^2-2m_0|\bar{\omega}|\left(\mu_l+\frac{s\Phi_{AC}}{\pi}-\frac{s}{2}\right)}{2m_0|\bar{\omega}|}\right]-\frac{1+s}{2}-\mathbb{B}_3^s,
\end{align}
where the operator $\mathbb{B}_3^s$ is defined as
\begin{align}\nonumber\label{oper31}
\mathbb{B}_3^s\varphi_s\equiv&\frac{1}{4m_0|\bar{\omega}|}\left[-\frac{d^2}{d\rho^2}-\frac{1}{\rho}\frac{d}{d\rho}+m_0^2\bar{\omega}^2\rho^2+\frac{\Gamma_s^2}{\rho^2}+m_0^2-2m_0|\bar{\omega}|\Gamma_s-2m_0|\bar{\omega}|\right]\varphi_s\\
=&\frac{1}{4m_0|\bar{\omega}|}\left[\mu B -E\right]^2\varphi_s.
\end{align}
Now, if we introduce the operator $\mathbb{B'}_3^s$ defined by
\begin{equation}\label{BEP}
\mathbb{B'}_3^s=\mathbb{B}_3^s+\frac{1}{4\pi|\bar{\omega}|}\left[|\bar{\omega}|\pi-m_0\pi+2|\bar{\omega}|m_l\pi+2|\bar{\omega}|\Phi_{AC}\right],
\end{equation}
then we can show that both operators $\mathbb{B}_{\pm(1)}^s$ (for $s=1$) and $\mathbb{B'}_3^s$, as well as $\mathbb{B}_{\pm(2)}^s$ (for $s=-1$) and $\mathbb{B'}_3^s$ close the $su(1,1)$ Lie algebra of equation (\ref{comm}) of Appendix, i.e.,
\begin{equation}\label{com1}
\left[\mathbb{B'}_3^s, \mathbb{B}_{\pm(1,2)}^s\right]=\pm\mathbb{B}_{\pm(1,2)}^s, \hspace{0.5cm} \left[\mathbb{B}_{-(1,2)}^s,\mathbb{B}_{+(1,2)}^s\right]=2\mathbb{B'}_3^s.
\end{equation}
Notice that these operators directly depend on the magnetic field strength, the AC frequency and the AC phase.

A directly computation shows that the action of the Casimir operator $\mathit{C}_s^2$ on $\varphi_s$ for this algebra realization satisfies the eigenvalue equation
\begin{equation}\label{Cas1}
\mathit{C}_s^2\varphi_s=\frac{1}{4}\left[\Gamma_s^2-1\right]\varphi_s=k
(k-1)\varphi_s,
\end{equation}
where the last equality is due to the $su(1,1)$ theory of unitary representations (see Appendix). Therefore,  we can obtain the relationship between $\Gamma_s$ and the Bargmann quantum number $k$
\begin{equation}\label{numk}
k_s=\frac{1}{2}|\Gamma _s|+\frac{1}{2}.
\end{equation}
The other group number $n$ can be identified with the radial quantum number, i.e. $n=n_r$. By comparison of equations (\ref{BEP}) and (\ref{k0n}), and using the previous results (\ref{cmvar}) and (\ref{xi}) we obtain
\begin{equation}\label{Ener}
n+k_s=n_r+\frac{1}{2}|\Gamma_s|+\frac{1}{2}=\frac{1}{4m_0|\bar{\omega}|}\left[\mu B -E\right]^2+\frac{1}{4\pi|\bar{\omega}|}\left[|\bar{\omega}|\pi-m_0\pi+2|\bar{\omega}|m_l\pi+2|\bar{\omega}|\Phi_{AC}\right],
\end{equation}
with $n_r=0,1,2,...$

Therefore for both indexes $(\pm(1),\pm(2))$, the energy spectrum for particle ($+$) and antiparticle ($-$) can be computed from equation (\ref{Ener}) to obtain
\begin{equation}\label{esMs}
E_{\pm(1,2)}^{n_s,\mu_l}=\mu\beta\pm\sqrt{m_0^2+4m_0|\bar{\omega}|\left[n_s+\frac{|\Gamma_s|}{2}-\frac{\Gamma_s}{2}\right]},\hspace{0.5cm} n_s=n_r+\frac{1-s}{2}.
\end{equation}
In this expression we observe that the Aharonov-Casher phase $\Phi_{AC}$ has periodicity $\Phi_0=\pm 2\pi$ and therefore $E_{\pm(1,2)}^{n_s,\mu_l}\left(\Phi_{AC}\pm 2\pi\right)=E_{\pm(1,2)}^{n_s,\mu_{l+1}}\left(\Phi_{AC}\right)$. Thus, the energy spectrum is a periodic function with periodicity $\pm 2\pi$.

Now, in order to obtain the explicit form of the radial wave function, in equation (\ref{secondAC}) we define $\varphi_s(\rho)=\frac{1}{\sqrt{\rho}}\Psi(\rho)$. Thus, the change $m_0|\bar{\omega}|\rho^2 \rightarrow r^2$ leads to
\begin{equation}\label{second2AC2}
\left(\frac{d^2}{dr^2}+\frac{\frac{1}{4}-\Gamma_s^2}{r^2}+\frac{E_s}{m_o|\bar{\omega}|}-r^2
 \right)\Psi(\rho)=0.
\end{equation}
From equation (\ref{second2AC2}), the radial wave function has the form \cite{LEB}
\begin{equation}\label{second3}
\Psi(\rho)=\left[\frac{2\Gamma\left(n_r+1\right)}{\Gamma_s\left(n_r+|\Gamma_s|+1\right)}\right]^{1/2}e^{\frac{-m_0|\bar{\omega}|\rho^2}{2}}
\left(m_0|\bar{\omega}|\right)^{\frac{|\Gamma_s|+\frac{1}{2}}{2}}\rho^{|\Gamma_s|+\frac{1}{2}} L_{n_r}^{|\Gamma_s|}(m_0|\bar{\omega}|\rho^2),
\end{equation}
where the normalization coefficient $N_n$ was computed from  the orthogonality of the Laguerre polynomials. Thus, the Sturmian basis for the DO interacting with the AC system in the Ms in terms of the group numbers $n$, $k_{s}=\frac{1}{2}|\Gamma_s|+\frac{1}{2}$ are
\begin{equation}
\varphi(\rho)_{n_r,\Gamma_s}=\left[\frac{2\Gamma\left(n_r+1\right)}{\Gamma\left(n_r+|\Gamma_s|+1\right)}\right]^{\frac{1}{2}}
e^{\frac{-m_0|\bar{\omega}|\rho^2}{2}}\left(m_0|\bar{\omega}|\right)^{\frac{|\Gamma_s|+\frac{1}{2}}{2}}\rho^{|\Gamma_s|}L_{n_r}^{|\Gamma_s|}(m_0|\bar{\omega}|\rho^2),
\end{equation}
or
\begin{equation}\label{sturm1}
\varphi(\rho)_{n_r,k_s}=\left[\frac{2\Gamma\left(n_r+1\right)}{\Gamma\left(n+2k_s\right)}\right]^{\frac{1}{2}}e^{\frac{-m_0|\bar{\omega}|\rho^2}{2}}
\left(m_0|\bar{\omega}|\right)^{\frac{2k_s-\frac{1}{2}}{2}}\rho^{2k_s-1} L_{n}^{2k_s-1}(m_0|\bar{\omega}|\rho^2).
\end{equation}
According to the equations (\ref{k+n}) and (\ref{k-n}), the action of the operators $\mathbb{B}_{\pm(1,2)}^s$  on this function is given by

\begin{equation}
\mathbb{B}_{\pm(1,2)}^s\varphi(\rho)_{n_r,k_s}=Q_{\pm(1,2)}^{({n_r,k_s})}\varphi(\rho)_{n_r\pm1,k_s},
\end{equation}
where
\begin{equation}
 Q_{\pm(1,2)}^{({n_r,k_s})}=\left[\left[\left(n_r+k_s\right)\mp\left(k_s-1\right)\right]\left[\left(n_r+k_s\right)\pm\left(k_s-1\right)\pm1\right]\right]^{\frac{1}{2}}.
\end{equation}

\subsection{$SU(1,1)$ radial coherent states and their time evolution}

In this Section we will compute the $SU(1,1)$ Perelomov coherent states $|\zeta\rangle=D\left(\xi\right)|k,0\rangle$ for the radial functions by using the Sturmian basis of equation (\ref{sturm1}), where $D\left(\xi\right)$ is the displacement operator and $|k,0\rangle$ the lowest normalized state (see Appendix). Hence, from the normal form of the displacement operator we obtain
\begin{equation}
|\zeta\rangle=D\left(\xi\right)|k_s,0\rangle=\exp(\zeta \mathbb{B}_{+(1,2)})\exp(\eta \mathbb{B'}_3^s)\exp(-\zeta^*\mathbb{B}_{-(1,2)})|k_s,0\rangle.
\end{equation}
The action of these exponentials on a state $|k_s,0\rangle$ explicitly is
\begin{align}
\sum_{j=0}^{\infty}\frac{\left(-\zeta^*\mathbb{B}_{-(1,2)}\right)^j}{j!}||\Gamma_s|,n\rangle=&\sum_{j=0}^{n}\frac{\left(-\zeta^*\right)^j}{j!}\left(\frac{n_r!\left(|\Gamma_s|+n_r\right)!}{\left(n_r-j\right)!\left(|\Gamma_s|+n_r-1-j\right)!}\right)^{1/2}||\Gamma_s|,n_r-j\rangle,\\
\sum_{j=0}^{\infty}\frac{\left(\eta \mathbb{B'}_3^s\right)^j}{j!}||\Gamma_s|,n_r\rangle=&\sum_{j=0}^{\infty}\frac{\left(\eta\left(\frac{1}{2}|\Gamma_s|+\frac{1}{2}+n_r\right)\right)^j}{j!}||\Gamma_s|,n_r\rangle,\\
\sum_{j=0}^{\infty}\frac{\left(\zeta \mathbb{B}_{+(1,2)}\right)^j}{j!}||\Gamma_s|,n_r\rangle=&\sum_{j=0}^{\infty}\frac{\zeta^j}{j!}\left(\frac{\left(n_r+j\right)!\left(|\Gamma_s|+n_r+j\right)!}{n!\left(|\Gamma_s|+n_r\right)!}\right)^{1/2}||\Gamma_s|,n_r-j\rangle.
\end{align}
From these results we obtain
\begin{equation}\label{ECAC1}
\varphi_s(\rho,\xi)=\left[\frac{2\left(1-|\xi|^2\right)^{2k_s}}{\Gamma\left(2k_s\right)}\right]^{\frac{1}{2}}\rho^{2k_s-1}e^{\frac{-m_0|\bar{\omega}|\rho^2}{2}}\left(m_0|\bar{\omega}|\right)^{\frac{2k_s-\frac{1}{2}}{2}}\sum_{n=0}^\infty\xi^nL_{n_r}^{2k_s-1}\left(m_0|\bar{\omega}|\rho^2\right).
\end{equation}
The sum of this equation can be calculated from the Laguerre polynomials generating function
\begin{equation}\label{SUML}
\sum_{s=0}^\infty L_n^\nu(x)y^n=\frac{e^{-xy/(1-y)}}{\left(1-y\right)^{y+1}}.
\end{equation}
Therefore, the $SU(1,1)$ radial coherent states for the DO including the AC effect $\varphi_s$ can be written as
\begin{equation}\label{estcoh1}
\varphi_s(\rho,\xi)=\left[\frac{2\left(1-|\xi|^2\right)^{2k_s}}{\Gamma\left(2k_s\right)} \frac{\left(m_0|\bar{\omega}|\right)^{2k_s-\frac{1}{2}}}{\left(1-\xi\right)^{4k_s}}\right]^{\frac{1}{2}}\rho^{2k_s-1}e^{\frac{m_0|\bar{\omega}|\rho^2}{2}\left(\frac{\xi+1}{\xi-1}\right)},
\end{equation}
or in terms of the physical quantum number $|\Gamma_s|$
\begin{equation}\label{estcoh2}
\varphi_s(\rho,\xi)=\left[\frac{2\left(1-|\xi|^2\right)^{|\Gamma_s|+1}}{\Gamma\left(|\Gamma_s|+1\right)} \frac{\left(m_0|\bar{\omega}|\right)^{|\Gamma_s|+\frac{1}{2}}}{\left(1-\xi\right)^{2|\Gamma_s|+2}}\right]^{\frac{1}{2}}\rho^{|\Gamma_s|+\frac{1}{2}}e^{\frac{m_0|\bar{\omega}|\rho^2}{2}\left(\frac{\xi+1}{\xi-1}\right)}.
\end{equation}

In this work we are only interested in computing the Perelomov coherent states. However, since the symmetry of our problem is that of the $SU(1,1)$ group, we can similarly obtain the Barut-Girardello coherent states. As can be seen in Ref. \cite{RosasBG}, these Barut-Girardello coherent states of the three-dimensional harmonic oscillator are expressed in terms of the Bessel function of the first kind.

Moreover, the equation (\ref{second2AC2}) can be written as
\begin{equation}\label{Hgam}
H_r\varphi_s=\xi_s\varphi_s,
\end{equation}
with
\begin{equation}\label{second4}
H_r=\left(-\rho^2\frac{d^2}{d\rho^2}+\frac{\Gamma_s^2-\frac{1}{4}}{\rho^2}+m_0^2\bar{\omega}^2\rho^2\right).
\end{equation}
Therefore, from equations (\ref{BEP}) and (\ref{Hgam}) we obtain
\begin{equation}\label{HAMAC}
H_r\varphi_s=4m_0|\bar{\omega}|\mathbb{B}_3^{'s}\varphi_s=\xi_s\varphi_s.
\end{equation}
From this expression, we get in an alternative way the energy spectrum for the DO including the AC effect  in the Ms
\begin{equation}
\xi_{\pm(1,2)}^{n_s,\mu_l}=\mu\beta\pm\sqrt{m_0^2+4m_0|\bar{\omega}|\left[n_s+\frac{|\Gamma_s|}{2}-\frac{\Gamma_s}{2}\right]}.
\end{equation}

Now, since the Hamiltonian is proportional to $\mathbb{B}_3^{'s}$, the time evolution of the coherent states of equation (\ref{estcoh2}) can be computed from the operator $\mathcal{U}(t)=e^{-iH_rt/\hbar}=e^{-i\gamma \mathbb{B}_3^{'s}t/\hbar}$ \cite{Cohen}. Thus,
\begin{equation}\label{PERET}
|\zeta(t)\rangle =\mathcal{U}(t)|\zeta\rangle=\mathcal{U}(t)D(\xi)\mathcal{U}^\dag(t)\mathcal{U}(t)|k_s,0\rangle.
\end{equation}
From the BCH formula it can be shown that the time evolution of the displacement operator $D(\xi)$ is due to the time evolution of the complex $\xi(t)=\xi e^{i\gamma t/\hbar}$ \cite{MSR1,MSR3}. Thus,
\begin{equation}
|\zeta(t)\rangle=e^{-i\gamma k_st/\hbar}e^{\zeta(-t)\mathbb{B}_{+(1,2)}}e^{\eta \mathbb{B}_3^{'s}}e^{-\zeta(-t)^*\mathbb{B}_{-(1,2)}}|k_s,0\rangle.
\end{equation}
Therefore, from these results, we can compute the temporal evolution of the coherent states of equation (\ref{estcoh2}) to obtain
\begin{equation}
\varphi_s^{k_s}(\rho)=\left[\frac{2\left(1-|\xi|^2\right)^{2k_s}\left(m_0\bar{\omega}\right)^{2k_s-\frac{1}{2}}}{\Gamma\left(2k_s\right)
\left(1-\xi{e}^{4im_0\bar{\omega}\tau/\hbar}\right)^{4k_s}}\right]^{\frac{1}{2}}\rho^{2k_s-1}
  e^{-4im_0|\bar{\omega}|\left(k_s\right)}e^{\frac{m_0|\bar{\omega}|\rho^2}{2}\left(\frac{\xi{e}^{4im_0|\bar{\omega}|\tau/\hbar}+1}{\xi{e}^{4im_0|\bar{\omega}|\tau/\hbar}-1}\right)}.
\end{equation}

\section{The DO and the AC effect in the Css}

The DO interacting with topological defects has been widely studied from different approaches\cite {RRS}. In one of these works, Carvalho et al. \cite{Carv} obtained the energy levels and eigenfunctions of the DO in the presence of three defects, the cosmic string, the magnetic cosmic string, and the cosmic dislocation. In Ref. \cite{Bakkee}, the authors consider the Css background to study the influence of non-inertial effects on the Dirac oscillator.

The metric tensor for the Css in cylindrical coordinates is defined by the line element
 \begin{equation}
ds^2=c^2dt^2-d\rho^2-\eta^2\rho^2d\theta^2-dz^2,
\end{equation}
where the coordinates $(t,z)\in(-\infty,\infty)$, the angular variable $\theta\in[0,2\pi]$, $\eta=1-\frac{4\bar{m}G}{c^2}$ is  a parameter related to the deficit angle and $\bar{m}$ is the linear mass density of the cosmic string \cite{Fur,Carv,EFM,ERB}. In this line element the geometry has a conical singularity  $R_{\rho, \theta}^{\rho, \theta}$ that gives rise to the curvature centered on the cosmic string axis ($z$-axis), in other places the curvature  is null.

Thus, the Dirac equation for this problem can be written as
\begin{equation}\label{ECUDCS}
\left[i\gamma^{\mu}(x)\left(\nabla_{\mu}(x)+m_0\omega\rho\gamma^0\delta_{\mu}^{\rho}\right)+\frac{\mu}{2}\sigma^{\mu\nu}(x)F_{\mu\nu}-m_0\right]\Psi(t,\textbf{r})=0,
\end{equation}
with $\mu, \nu=0,1,2,3 $ and where $\gamma^{\mu}(x)$ are the generalized Dirac matrices. These matrices satisfy the relation $\gamma^{\mu}\gamma^{\nu}+\gamma^{\nu}\gamma^{\mu}=2g^{\mu\nu}(x)$ ($g^{\mu\nu}(x)$ is the metric tensor). In equation (\ref{ECUDCS}), $\nabla_{\mu}(x) =\partial_{\mu}+\Gamma_\mu(x)$ is the covariant derivative, with $\Gamma_\mu(x)$ the spinor affine connection and $\delta_{\mu}^{\rho}$ is the Kronecker delta. The $\gamma^{\mu}(x)$ matrices can be written in terms of the standard Dirac matrices $\gamma^{a}$ in Ms as
\begin{equation}
\gamma^{\mu}(x)=e_{(a)}^{\mu}(x)\gamma^a,
\end{equation}
where the tetrad basis $e_{(a)}^{\mu}(x)$ satisfies the relations
\begin{equation}
e_{a}^{(\mu)}(x)e_{b}^{(\nu)}(x)\eta^{ab}=g^{\mu\nu}(x), \hspace{0.5 cm} e_{\mu}^{(a)}(x)e_{b}^{(\mu)}(x)=\delta_{b}^{a},\hspace{0.5 cm} e_{a}^{(\mu)}(x)e_{\nu}^{(a)}(x)=\delta_{\nu}^{\mu}.
\end{equation}
The  tetrad basis $e_{(\mu)}^{a}(x)$ and $e_{(a)}^{\mu}(x)$ in the Css are explicitly written as \cite{Carv}
\begin{equation}\label{matteCS}
e_{(a)}^{\mu}=\begin{pmatrix}
1 & 0 & 0 & 0 \\
0 & \cos\phi & \sin\phi & 0\\
0 & -\frac{\sin{\phi}}{\eta\rho} & \frac{\cos{\phi}}{\eta\rho} & 0\\
0 & 0 & 0 & 1
\end{pmatrix}, \hspace{0.5 cm}
e_{\mu}^{(a)}=\begin{pmatrix}
1 & 0 & 0 & 0 \\
0 & \cos\phi & -\eta\rho\sin\phi & 0\\
0 & \sin{\phi} & \eta\rho\cos{\phi} & 0\\
0 & 0 & 0 & 1
\end{pmatrix}.
\end{equation}
Thus, for this representation the Dirac matrices $\gamma^{\mu}(x)$ obey the relations
\begin{align}\nonumber
\gamma^{0}(x)=&\gamma^{t}, \hspace{0.5cm} \gamma^{1}(x)=\gamma^{\rho},\hspace{0.5cm} \gamma^{3}(x)=\gamma^{z},  \hspace{0.5cm}\gamma^{\rho}=\cos\phi\gamma^{1}+\sin\phi\gamma^{2},\\\label{Gamas}
\gamma^{2}=&\frac{\gamma^{\phi}}{\eta\rho}, \hspace{0.5cm} \gamma^{\phi}=-\sin\phi\gamma^{1}+\cos\phi\gamma^{2}.
\end{align}
Therefore, in terms of tetrad basis the equation (\ref{ECUDCS}) can be written as
\begin{equation}\label{ECUDCS2}
\left[ie_{a}^{\mu}(x)\gamma^a\left(\nabla_{\mu}(x)+m_0\omega\rho\gamma^0\delta_{\mu}^{\rho}\right)+\frac{\mu}{2}e_{a}^{\mu}(x)e_{b}^{\nu}(x)\sigma^{ab}F_{\mu\nu}-m_0\right]\Psi(t,\textbf{r})=0, \end{equation}
where the spinor affine connection is given by $\Gamma_{\mu}(x)=\frac{i}{4}\omega_{\mu{ab}}(x)\sigma^{ab}$, and $\omega_{\mu{ab}}(x)$ is the spin connection. With all these results, the Dirac equation for the DO including the AC effect in the Css can be written as \cite{RRS}
\begin{equation}\label{ARA-3}
\left[i\sigma_1\left(\partial_\rho-\frac{\left(1-\eta\right)}{2\eta\rho}+\sigma_3\left(mo\bar{\omega}\rho-\frac{s\Phi_{AC}}{\eta\pi\rho}\right)\right)+\sigma_2\left(\frac{i\partial_\theta}{\eta\rho}+\frac{\sigma_3}{2\eta\rho}\right)+\mu{B}+m_0\sigma_3-i\partial_0\right]\psi(t,\rho,\theta)=0,
\end{equation}
where $\psi(t,\rho,\theta)\equiv U^{-1}(\theta)\Psi(t,\rho, \theta)$, with $U(\theta)=e^{-\frac{i\theta\alpha_3}{2}}$. In particular, the differential equation for the radial part of this problem in the Css is \cite{RRS}

\begin{equation}\label{secondCS}
\left[\frac{d^2}{d\rho^2}+\frac{1}{\rho}\frac{d}{d\rho}-\frac{\gamma_s^2}{\eta^2\rho^2}-m_0^2\bar{\omega}^2\rho^2+E_s\right]\bar{\varphi}_s\left(\rho \right)=0,\hspace{0.5cm}
\end{equation}
where
\begin{equation}\label{cmvar2}
\gamma_s\equiv \mu_l+\frac{s\Phi_{AC}}{\pi}-\frac{s\eta}{2},\hspace{0.5cm} \bar{E}_s\equiv \left(\frac{\mu B}{\eta}-E\right)^2-m_0^2+\frac{2m_0\bar{\omega}\gamma_s}{\eta}+2sm_0\bar{\omega}.
\end{equation}

\section{Algebraic solution in the Css}

In this Section, we shall construct the $su(1,1)$ algebra generators for the DO with the AC effect in the Css \cite{MSR1,MSR2,Schro,Schro1,Schro2,Inf}. To this end, we will proceed as in Section $3$. Thus, we first express equation (\ref{secondCS}) as
\begin{equation}
\left[\mathbb{T}_-^s-1\right]\mathbb{T}_+^s=\frac{1}{4}\left[\left(\frac{\bar{E}_s}{2m_0|\bar{\omega}|}+1\right)^2-\frac{\gamma_s^2}{\eta^2}\right],
\end{equation}
where the Schr\"odinger operators $\mathbb{T}_\pm$ are given by
\begin{equation}\label{ED-COM2}
\mathbb{T}_\mp^s=\frac{1}{2}\left[\pm\rho\frac{d}{d\rho}+m_0|\bar{\omega}|\rho^2-\frac{1}{2m_0|\bar{\omega}|}\left[\left(\frac{\mu{B}}{\eta}-E_{cs}\right)^2-m_0^2+\frac{2m_0|\bar{\omega}|\gamma_s}{\eta}+2sm_0|\bar{\omega}|\right]+s \right].
\end{equation}
Applying the Schr\"odinger factorization method to equation (\ref{secondCS}), we obtain the following pair of operators
\begin{align}\label{OPSCHACS}
\mathbb{S}_{\pm(1)}^s=&\frac{1}{2}\left[\mp\rho\frac{d}{d\rho}+m_0|\bar{\omega}|\rho^2-\frac{m_0^2-\frac{2m_0|\bar{\omega}|}{\eta}\left(\mu_l+\frac{s\Phi_{AC}}{\pi}-\frac{s\eta}{2}\right)}{2m_0|\bar{\omega}|}\right]+\frac{1-s}{2}-\mathbb{T}_3^s,\\
\mathbb{S}_{\pm(2)}^s=&\frac{1}{2}\left[\mp\rho\frac{d}{d\rho}+m_0|\bar{\omega}|\rho^2-\frac{m_0^2-\frac{2m_0|\bar{\omega}|}{\eta}\left(\mu_l+\frac{s\Phi_{AC}}{\pi}-\frac{s\eta}{2}\right)}{2m_0|\bar{\omega}|}\right]-\frac{1+s}{2}-\mathbb{T}_3^s,
\end{align}
with $\mathbb{T}_3^s$ explicitly given by
\begin{align}\nonumber\label{oper3}
\mathbb{T}_3^s\chi_s\equiv&\frac{1}{4m_0|\bar{\omega}|}\left[-\frac{d^2}{d\rho^2}-\frac{1}{\rho}\frac{d}{d\rho}+m_0^2\bar{\omega}^2\rho^2+\frac{\gamma_s^2}{\eta^2\rho^2}+m_0^2-\frac{2m_0|\bar{\omega}|\gamma_s}{\eta}-2m_0|\bar{\omega}|\right]\chi_s\\
=&\frac{1}{4m_0|\bar{\omega}|}\left[\frac{\mu B}{\eta} -E_{cs}\right]^2\chi_s.
\end{align}
Then, we can show that both operators $\mathbb{S}_{\pm(1)}^s$ (for $s=1$) and $\mathbb{T'}_3^s$ as well as $\mathbb{S}_{\pm(2)}^s$ (for $s=-1$) and $\mathbb{T'}_3^s$ close the $su(1,1)$ Lie algebra of equation (\ref{comm}), i.e..
\begin{equation}\label{com2}
\left[\mathbb{T'}_3^s, \mathbb{T}_{\pm(1,2)}^s\right]=\pm\mathbb{T}_{\pm(1,2)}^s, \hspace{0.5cm} \left[\mathbb{T}_{-(1,2)}^s,\mathbb{T}_{+(1,2)}^s\right]=2\mathbb{T'}_3^s,
\end{equation}
where $\mathbb{T'}_3^s$ is expressed in terms of $\mathbb{T}_3^s$ as
\begin{equation}
\mathbb{T'}_3^s=\mathbb{T}_3^s+\frac{1}{4\pi\eta|\bar{\omega}|}\left[|\bar{\omega}|\eta\pi-m_0\eta\pi+2|\bar{\omega}|m_l\pi+2|\bar{\omega}|\Phi_{AC}\right].
\end{equation}

Therefore, in this case, the Casimir operator $\mathit{C}_{cs}^2$ and the quantum number $k_{cs}$ are given by
\begin{equation}\label{Cas}
\mathit{C}_{cs}^2\chi_s=\frac{1}{4}\left[\frac{\gamma_s^2}{\eta^2}-1\right]\chi_{cs}=k
(k-1)\chi_{cs},\hspace{1.0cm} k_{cs}=\frac{1}{2}\left|\frac{\gamma _s}{\eta}\right|+\frac{1}{2}.
\end{equation}
Thus, the energy spectrum for particle ($+$) and antiparticle ($-$) respectively for the DO including the AC effect in the Css is written as
\begin{equation}
E_{cs\pm(1,2)}^{n_s,\mu_l}=\frac{\mu\beta}{\eta}\pm\sqrt{m_0^2+4m_0|\bar{\omega}|\left[n_s+\frac{|\gamma_s|}{2\eta}-\frac{\gamma_s}{2\eta}\right]}.
\end{equation}
In this expression we observe that the Aharonov-Casher phase $\Phi_{AC}$ has periodicity $\Phi_0=\pm 2\pi$ and therefore $E_{n_s,\mu_l}\left(\Phi_{AC}\pm 2\pi\right)=E_{n_s,\mu_l+1}\left(\Phi_{AC}\right)$. Thus, the energy spectrum is a periodic function with periodicity $\pm 2\pi$. If we compare this spectrum with that calculated in equation (\ref{esMs}), we observe that the curvature of the Css slightly changes the pattern of the energy spectrum oscillations.

Thus, the radial wave functions are given by
\begin{equation}\label{second31}
\Psi_{cs}(\rho)=\left[\frac{2\Gamma\left(n_r+1\right)}{\Gamma_s\left(n_r+\left|\frac{\gamma_s}{\eta}\right|+1\right)}\right]^{1/2}e^{\frac{-m_0|\bar{\omega}|\rho^2}{2}}
\left(m_0|\bar{\omega}|\right)^{\frac{\left|\frac{\gamma_s}{\eta}\right|+\frac{1}{2}}{2}}\rho^{\left|\frac{\gamma_s}{\eta}\right|+\frac{1}{2}} L_{n_r}^{\left|\frac{\gamma_s}{\eta}
\right|}(m_0|\bar{\omega}|\rho^2),
\end{equation}
Moreover, the Sturmian basis for this case are
\begin{equation}\label{SB2}
\chi_{cs}(\rho)_{n_r,\gamma_s}=\left[\frac{2\Gamma\left(n_r+1\right)}{\Gamma\left(n_r+\left|\frac{\gamma_s}{\eta}\right|+1\right)}\right]^{\frac{1}{2}}
e^{\frac{-m_0|\bar{\omega}|\rho^2}{2}}\left(m_0|\bar{\omega}|\right)^{\frac{\left|\frac{\gamma_s}{\eta}\right|+\frac{1}{2}}{2}}\rho^{\left|\frac{\gamma_s}{\eta}\right|}L_{n_r}^{\left|\frac{\gamma_s}{\eta}\right|}(m_0|\bar{\omega}|\rho^2),
\end{equation}
or
\begin{equation}\label{sturm2}
\chi_{cs}(\rho)_{n_r,k_{cs}}=\left[\frac{2\Gamma\left(n_r+1\right)}{\Gamma\left(n+2k_{cs}\right)}\right]^{\frac{1}{2}}e^{\frac{-m_0|\bar{\omega}|\rho^2}{2}}
\left(m_0|\bar{\omega}|\right)^{k_{cs}+\frac{1}{2}}\rho^{2k_{cs}-1} L_{n}^{2k_{cs}-1}(m_0|\bar{\omega}|\rho^2).
\end{equation}

\subsection{$SU(1,1)$ radial coherent states and their time evolution}

Similarly to the procedure followed in Section $3.1$, we can compute the radial coherent states for the Sturmian basis of equation (\ref{sturm2}). Thus, from the definition of the $SU(1,1)$ Perelomov coherent states $|\zeta\rangle=D\left(\xi\right)|k,0\rangle$ we obtain
\begin{equation}\label{ECAC}
\chi_{cs}(\rho,\xi)=\left[\frac{2\left(1-|\xi|^2\right)^{2k_{cs}}}{\Gamma\left(2k_{cs}\right)}\right]^{\frac{1}{2}}\rho^{2k_{cs}-1}e^{\frac{-m_0|\bar{\omega}|\rho^2}{2}}\left(m_0|\bar{\omega}|\right)^{\frac{2k_{cs}-\frac{1}{2}}{2}}\sum_{n=0}^\infty\xi^nL_{n_r}^{2k_{cs}-1}\left(m_0|\bar{\omega}|\rho^2\right).
\end{equation}
Also, the sum in this expression can be calculated from the Laguerre polynomials given in equation (\ref{SUML}). Therefore, the radial coherent states can be written as
\begin{equation}\label{estcoh21}
\chi_{cs}(\rho,\xi)=\left[\frac{2\left(1-|\xi|^2\right)^{2k_{cs}}}{\Gamma\left(2k_{cs}\right)} \frac{\left(m_0|\bar{\omega}|\right)^{2k_{cs}-\frac{1}{2}}}{\left(1-\xi\right)^{4k_{cs}}}\right]^{\frac{1}{2}}\rho^{2k_{cs}-1}e^{\frac{m_0|\bar{\omega}|\rho^2}{2}\left(\frac{\xi+1}{\xi-1}\right)}.
\end{equation}
The $SU(1,1)$ radial coherent states for the DO including the AC effect in the Css in terms of the physical quantum number $|\gamma_s|$ are given by
\begin{equation}\label{estcoh2css}
\chi_{cs}(\rho,\xi)=\left[\frac{2\left(1-|\xi|^2\right)^{\left|\frac{\gamma_s}{\eta}\right|+1}}{\Gamma\left(\left|\frac{\gamma_s}{\eta}\right|+1\right)} \frac{\left(m_0|\bar{\omega}|\right)^{\left|\frac{\gamma_s}{\eta}\right|+\frac{1}{2}}}{\left(1-\xi\right)^{2\left|\frac{\gamma_s}{\eta}\right|+2}}\right]^{\frac{1}{2}}\rho^{\left|\frac{\gamma_s}{\eta}\right|+\frac{1}{2}}e^{\frac{m_0|\bar{\omega}|\rho^2}{2}\left(\frac{\xi+1}{\xi-1}\right)}.
\end{equation}

On the other hand, the differential equation (\ref{secondCS}) can be written as
\begin{equation}\label{second41}
H_{(cs)r}\chi_s=\left(-\rho^2\frac{d^2}{d\rho^2}+\frac{\Gamma_s^2-\frac{1}{4}}{\rho^2}+m_0^2\bar{\omega}^2\rho^2\right)\chi_s=4m_0|\bar{\omega}|\mathbb{T}_3^s\varphi_s=E_{cs}\chi_s.
\end{equation}
From this equation, we obtain again the energy spectrum for the DO including the AC effect in the Css
\begin{equation}\label{PCNCS}
E_{\pm(1,2)cs}^{n_s,\mu_l}=\frac{\mu\beta}{\eta}\pm\sqrt{m_0^2+4m_0|\bar{\omega}|\left[n_s+\frac{|\gamma_s|}{2\eta}-\frac{\gamma_s}{2\eta}\right]}.
\end{equation}

Now, from our results of Section $3.1$, we obtain that the temporal evolution of the states (\ref{PCNCS}) in terms of $k_{cs}$ are
\begin{equation}
\varphi_s^{k_s}(\rho)=\left[\frac{2\left(1-|\xi|^2\right)^{2k_{cs}}\left(m_0|\bar{\omega}|\right)^{2k_{cs}-\frac{1}{2}}}{\Gamma\left(2k_s\right)
\left(1-\xi{e}^{4im_0|\bar{\omega}|\tau/\hbar}\right)^{4k_{cs}}}\right]^{\frac{1}{2}}\rho^{2k_{cs}-1}e^{-4im_0|\bar{\omega}|\left(k_{cs}
  \right)}e^{\frac{m_0|\bar{\omega}|\rho^2}{2}\left(\frac{\xi{e}^{4im_0|\bar{\omega}|\tau/\hbar}+1}{\xi{e}^{4im_0|\bar{\omega}|\tau/\hbar}-1}\right)}.
\end{equation}

\section{Matrix elements and the \emph{Sur}}

In this Section, we will obtain some matrix elements for the radial wave function of the DO including the AC effect in the Ms and the Css. Also, we will obtain some expectation values and the \emph{Sur} for the Perelomov coherent states of this problem. From equations (\ref{OPSCHAC1}) and (\ref{OPSCHAC2}), the following relationships can be obtained

\begin{align}\label{rocua}
\rho^2 &=\frac{1}{m_0|\bar{\omega}|}\left[\mathbb{B}_+^s+\mathbb{B}_-^s+2\mathbb{B}_3^s+\mathbb{G}-1\right],\\
\rho\frac{d}{d\rho}&=\mathbb{B}_-^s-\mathbb{B}_+^s-1,
\end{align}
where
\begin{equation}
\mathbb{G}=\frac{-m_0^2-2m_0|\bar{\omega}|\left[\mu_l+\frac{\Phi_{AC}}{\pi}-\frac{s}{2}\right]+1}{m_0|\bar{\omega}|}.
\end{equation}
The matrix elements of these expressions can be computed algebraically by using the equations (\ref{k+n})-(\ref{k0n}) and are given by
\begin{align}
\langle \Psi_m|\rho^2|\Psi_n\rangle=&\frac{1}{m_0|\bar{\omega}|}\Bigl[ \sqrt{\left(n_s+1\right)\left(|\Gamma_s|+n_s+1\right)}\delta_{m,n}+\sqrt{n_s\left(|\Gamma_s|+n_s\right)}\delta_{m,n}\\\nonumber
+&\left[|\Gamma_s|+2n_s+1\right]\delta_{m,n}+\mathbb{G}\Bigr],\\
\langle\Psi_m|\rho\frac{d}{d\rho}|\Psi_n\rangle=&\sqrt{n_s\left(|\Gamma_s|+n_s\right)}\delta_{m,n}-\sqrt{\left(n_s+1\right)\left(|\Gamma_s|+n_s+1\right)}\delta_{m,n}-1.
\end{align}

Now, by using the Baker-Campbell-Hausdorff identity
\begin{equation}
e^{-A}Be^{A}=B+\frac{1}{1!}\left[B,A\right]+\frac{1}{2!}\left[\left[B,A\right], A\right]+\frac{1}{3!}\left[\left[\left[B,A\right], A\right], A\right]+...,
\end{equation}
and the $su(1,1)$ commutation relations of equation (\ref{comm}), the following similarity transformations are obtained for the $su(1,1)$ Lie algebra generators
\begin{align}\label{sim1}
D^\dag(z)\mathbb{B}_+^sD(z)=&\frac{z^*}{|z|}\alpha\mathbb{B}_3^s+\beta\left(\mathbb{B}_+^s+\frac{z^*}{z}\mathbb{B}_-^s\right)+\mathbb{B}_+^s,\\\label{sim2}
D^\dag(z)\mathbb{B}_-^sD(z)=&\frac{z}{|z|}\alpha\mathbb{B}_3^s+\beta\left(\mathbb{B}_-^s+\frac{z}{z^*}\mathbb{B}_+^s\right)+\mathbb{B}_-^s,\\\label{sim3}
D^\dag(z)\mathbb{B}_3^sD(z)=&\left(2\beta+1\right)\mathbb{B}_3^s+\frac{{\alpha}z}{2|z|}\mathbb{B}_+^s+\frac{{\alpha}z^*}{2|z|}\mathbb{B}_-^s,
\end{align}
with $\alpha=\sinh(2|z|)$ and $\beta=\frac{1}{2}\cosh(2|z|-1)$. Thus, with the help of equations (\ref{sim1})-(\ref{sim3}), we obtain the following expectation values of the group generators $\mathbb{B}_+^s, \mathbb{B}_-^s$ and $\mathbb{B}_3^s$ for the Perelomov coherent states
\begin{align}
\langle \xi|\mathbb{B}_+^s|\xi\rangle=&\frac{z^*}{|z|}\sinh\left(2|z|\right)\left(\frac{1}{2}|\Gamma_s|+\frac{1}{2}\right),\\
\langle \xi|\mathbb{B}_-^s|\xi\rangle=&\frac{z}{|z|}\sinh\left(2|z|\right)\left(\frac{1}{2}|\Gamma_s|+\frac{1}{2}\right),\\
\langle \xi|\mathbb{B}_3^s|\xi\rangle=&\cosh\left(2|z|\right)\left(\frac{1}{2}|\Gamma_s|+\frac{1}{2}\right).
\end{align}

\subsection{Schr\"odinger uncertainty relationship}

Now we are going to study whether the coherent states calculated in this work are of minimal uncertainty or not. To do this, from the $SU(1,1)$ group ladder operators $\mathbb{B}_+^s$ and $\mathbb{B}_-^s$, we can define the operators $X$ and $Y$ as follows
\begin{equation}
X\equiv\mathbb{B}_+^s+\mathbb{B}_-^s,\hspace{0.5cm}Y\equiv i\left(\mathbb{B}_+^s-\mathbb{B}_-^s\right).
\end{equation}
The \emph{Sur} for these general operators states that
\begin{equation}
\left(\Delta X\right)^2\left(\Delta Y\right)^2\geq\langle F \rangle^2+\frac{1}{4}\langle C \rangle^2.\label{schrour}
\end{equation}
Here, $\langle C \rangle\equiv -i\langle|X,Y|\rangle$ and $\langle F \rangle\equiv\langle\frac{1}{2}\{X,Y\}+\langle X\rangle\langle Y\rangle\rangle$ is the quantum correlation of the operators $X$ and $Y$. Also, in this expression
\begin{equation}
(\Delta X)^{2}=\langle\xi|X^{2}|\xi\rangle-\langle\xi|X|\xi\rangle^{2}, \quad\quad\quad(\Delta Y)^{2}=\langle\xi|Y^{2}|\xi\rangle-\langle\xi|Y|\xi\rangle^{2}.
\end{equation}

By using the definition of the Perelomov coherent states (\ref{normal}) and the similarity transformations (\ref{sim1}) and (\ref{sim2}), we obtain
\begin{align}
\langle\xi|X^2|\xi\rangle=&\alpha^2\left(\frac{1}{2}|\Gamma_s|+\frac{1}{2}\right)^2\left(2+\frac{z^*}{z}+\frac{z}{z^*}\right)+2\left(\frac{1}{2}|\Gamma_s|+\frac{1}{2}\right)\left[\left(2+\frac{z^*}{z}+\frac{z}{z^*}\right)\left(\lambda^2+\lambda\right)+1\right],\\
\langle\xi|Y^2|\xi\rangle=&\alpha^2\left(\frac{1}{2}|\Gamma_s|+\frac{1}{2}\right)^2\left(2-\frac{z^*}{z}-\frac{z}{z^*}\right)+2\left(\frac{1}{2}|\Gamma_s|+\frac{1}{2}\right)\left[\left(2-\frac{z^*}{z}-\frac{z}{z^*}\right)\left(\lambda^2+\lambda\right)+1\right],\\
\langle\xi|X|\xi\rangle=&\frac{\alpha\left(\frac{1}{2}|\Gamma_s|+\frac{1}{2}\right)}{|z|}\left(z^*+z\right),\\
\langle\xi|Y|\xi\rangle=&\frac{i\alpha\left(\frac{1}{2}|\Gamma_s|+\frac{1}{2}\right)}{|z|}\left(z^*-z\right).
\end{align}
From these results the quadratic deviations of the $X$ and $Y$ operators can be computed, to obtain\cite{DOGu3}
\begin{align}
\left(\Delta X\right)^2=&2\left(\frac{1}{2}|\Gamma_s|+\frac{1}{2}\right)\left[\left(2+\frac{z^*}{z}+\frac{z}{z^*}\right)\left(\lambda^2+\lambda\right)+1\right],\\
\left(\Delta Y\right)^2=&2\left(\frac{1}{2}|\Gamma_s|+\frac{1}{2}\right)\left[\left(2-\frac{z^*}{z}-\frac{z}{z^*}\right)\left(\lambda^2+\lambda\right)+1\right],
\end{align}
and the product of these factors is
\begin{equation}\label{prodxy}
\left(\Delta X\right)^2\left(\Delta Y\right)^2= 4\left(\frac{1}{2}|\Gamma_s|+\frac{1}{2}\right)^2\left[\left(\lambda^2+\lambda\right)^2\left(4-\left(\frac{z^*}{z}+\frac{z}{z^*}\right)^2\right)+4\left(\lambda^2+\lambda\right)+1\right].
\end{equation}

On the other hand, the expectation values $\langle F \rangle$ and $\langle C \rangle$ evaluated in the Perelomov coherent state are given by
\begin{equation}
\langle\xi|F|\xi\rangle=2i\left(\frac{1}{2}|\Gamma_s|+\frac{1}{2}\right)(\lambda^2+\lambda)\left(\frac{z^{*}}{z}-
\frac{z}{z^{*}}\right),\label{fn}
\end{equation}
\begin{equation}
\langle\xi|C|\xi\rangle=4\left(\frac{1}{2}|\Gamma_s|+\frac{1}{2}\right)(2\lambda+1)\label{cn}.
\end{equation}
By substituting the results of equations (\ref{prodxy}), (\ref{fn}) and (\ref{cn}) into equation (\ref{schrour}), we conclude that the coherent states are of minimum uncertainty, according to the \emph{Sur}. This result is in full agreement to that previously reported in Ref. \cite{Perelomov}.

All the results obtained in this Section can be translated to the case where the DO is now in the Css by replacing $\Gamma_s$ by $\frac{\gamma_s}{\eta}$ and $\mathbb{B}_+^s, \mathbb{B}_-^s, \mathbb{B}_3^s$ by $\mathbb{T}_+^s, \mathbb{T}_-^s, \mathbb{T}_3^s$ respectively.

\section{Concluding remarks}

We have shown that the Dirac equation for the Dirac oscillator including the Aharonov-Casher effect coupled to an external electromagnetic field in the Ms and the Css can be solved from an algebraic approach. Furthermore, by using the Schr\"odinger factorization we were able to build the generators of this algebra which show a dependence of the magnetic field strength, the AC frequency $\omega_{AC}$ and the AC phase $\Phi_{AC}$. We generalized the radial Schr\"odinger factorization operators in order to close an $su(1,1)$ Lie algebra. Using the theory of unitary representations for this algebra we obtained the energy spectrum and the eigenfunctions which also depend on the magnetic field strength, $\omega_{AC}$ and $\Phi_{AC}$. We obtained the Sturmian basis of this Lie algebra, which allowed us to construct the Perelomov coherent states and their time evolution. Finally, we introduced two general operators in terms of the $su(1,1)$ Lie algebra generators to calculate some matrix elements. Also, we showed that the \emph{Sur} is minimal for these coherent states.

Ioffe et al. \cite{Ioffe} studied a Pauli Hamiltonian describing the interaction of a neutral spin-$1/2$ particle interacting with a magnetic field generated by an electric current-carrying straight wire. Once the separation of variables has been carried out, the one-dimensional matrix Hamiltonian is analyzed from the point of view of supersymmetric quantum mechanics and constructed intertwining operators that depend on the matrix superpotentials. However, in our work we applied the Schr\"odinger factorization to the uncoupled radial equations, while in Ref. \cite{Ioffe} the authors used a matrix factorization. On the other hand, in Ref. \cite{Schul}, the Darboux transformations are used for the construction of zero-energy solutions to the massless Dirac equation with a radially symmetric, diagonal matrix potential. Therefore, we conclude that the essence of the methods followed in the Refs. \cite{Ioffe,Schul} are different to that we used in the present work. However, we can try to apply these algebraic methods to the problem we studied in the present paper in future works.

On the other hand, in Ref. \cite{RosasBG}, the authors factorized the radial part of the spherical oscillator Hamiltonian in four different ways. They showed that the factorization operators can be used to construct the generators of the Lie algebras $su(1,1)$ and $su(2)$. The factorization method used in that work to obtain their algebra generators is different to that used in the present work. However, since the problem studied in our work mathematically is reduced to the three-dimensional harmonic oscillator Hamiltonian (equation (\ref{secondAC})), another kinds of factorizations can be applied to analyze other aspects of the problem treated in the present paper.

To our knowledge, this is the first time where the Dirac oscillator including the influence of the Aharonov-Casher effect in the cosmic string background has been treated by an $su(1,1)$ algebraic approach. \\

\section*{Acknowledgments}

This work was partially supported by SNI-M\'exico, COFAA-IPN, EDI-IPN, EDD-IPN, SIP-IPN project numbers $20211834$ and $20210734$.

\section*{Appendix. The SU(1,1) Group and its coherent states}
\renewcommand{\theequation}{A\arabic{equation}}
\setcounter{equation}{0}

The $su(1,1)$ Lie algebra is defined by the following relations \cite{Barut,Perelomov}
\begin{eqnarray}
[K_1,K_2]=-iK_3, \quad\quad [K_2,K_3]=iK_1, \quad\quad [K_3,K_1]=iK_2.
\end{eqnarray}
In these expressions, the operators $K_i$, $i=1,2,3$ are the generators of the Lie algebra. Introducing the standard raising and lowering operators $K_{\pm}$
\begin{equation}
K_{\pm}=K_1\pm iK_2,
\end{equation}
the commutation relations can be written as
\begin{eqnarray}
[K_{3},K_{\pm}]=\pm K_{\pm},\quad\quad [K_{-},K_{+}]=2K_{3}.\label{comm}
\end{eqnarray}

The quadratic Casimir operator $C^2$ for this algebra, which commutes with all the algebra generators, has the form
\begin{equation}
C^2=K_3^2-K_1^2-K_2^2=K_3^2-\frac{1}{2}\left(K_+K_-+K_-K_+\right), \quad\quad [C^2,K_i]=0.
\end{equation}
The $su(1,1)$ algebra has different series of unitary representations, but here we consider
only the discrete series. A basis for an irreducible representation is given by the set
$\{|k,n\rangle, n=0,1,2,...\}$, where $k$ is the representation index, also called the Bargmann index.
The action of these operators on this basis states is
\begin{equation}
K_{+}|k,n\rangle=\sqrt{(n+1)(2k+n)}|k,n+1\rangle,\label{k+n}
\end{equation}
\begin{equation}
K_{-}|k,n\rangle=\sqrt{n(2k+n-1)}|k,n-1\rangle,\label{k-n}
\end{equation}
\begin{equation}
K_{3}|k,n\rangle=(k+n)|k,n\rangle,\label{k0n}
\end{equation}
where $|k,0\rangle$ is the lowest normalized state. The value of the Casimir operator $C^2$ in the invariant space $\{|k,n\rangle\}$ is equal to $k(k-1)$.

Now the $SU(1,1)$ Perelomov coherent states $|\zeta\rangle$ are defined in terms of the displacement operator $D(\xi)=\exp(\xi K_{+}-\xi^{*}K_{-})$ as \cite{Perelomov2,Perelomov,Gilmore2}
\begin{equation}
|\zeta\rangle=D(\xi)|k,0\rangle.\label{defPCS}
\end{equation}
Since the ladder operators $K_{\pm}$ satisfy the properties $K^{\dag}_{+}=K_{-}$ and $K^{\dag}_{-}=K_{+}$, it can be shown that
the displacement operator possesses the property
\begin{equation}
D^{\dag}(\xi)=\exp(\xi^{*} K_{-}-\xi K_{+})=D(-\xi).\label{dis}
\end{equation}

A more useful representation of the displacement operator $D(\xi)$ is given by the so-called normal form of this operator
\begin{equation}
D(\xi)=\exp(\zeta K_{+})\exp(\eta K_{3})\exp(-\zeta^*K_{-})\label{normal},
\end{equation}
where $\xi=-\frac{1}{2}\tau e^{-i\varphi}$, $\zeta=-\tanh
(\frac{1}{2}\tau)e^{-i\varphi}$ and $\eta=-2\ln \cosh
|\xi|=\ln(1-|\zeta|^2)$ \cite{GER}. This normal form of the displacement
operator and equations (\ref{k+n})-(\ref{k0n}) can be used to obtain the following expression of the Perelomov coherent states \cite{Perelomov}
\begin{equation}
|\zeta\rangle=(1-|\xi|^2)^k\sum_{s=0}^\infty\sqrt{\frac{\Gamma(n+2k)}{s!\Gamma(2k)}}\xi^s|k,s\rangle.\label{PCN}
\end{equation}

\end{document}